\documentstyle[ltwol]{article}
\input{psfig}

\def\be{\begin{equation}}
\def\ee{\end{equation}}
\def\bea{\begin{eqnarray}}
\def\eea{\end{eqnarray}}

\begin{document}
\title{~ ~ ~ ~ ~ ~ ~ ~ ~ ~ ~ ~ ~ ~ ~ ~ ~ ~ ~ ~ ~ ~ ~ ~ ~ ~ ~ ~ ~ ~~~~~~~~~~~~~~~~~~~~~~~~~~~~~~~~~~~~~~~~~~~
IFT/16/2001 \newline\newline\newline
PHOTOPRODUCTION OF THE ISOLATED PHOTON AT HERA IN NLO QCD}
\author{M. KRAWCZYK, A. ZEMBRZUSKI}
\address{Institute of Theoretical Physics, Warsaw University,
ul. Ho\.za 69, 00-681 Warsaw, 
Poland\\E-mail: krawczyk@fuw.edu.pl,azem@fuw.edu.pl}   

\twocolumn[\maketitle\abstracts{ 
The NLO QCD calculation for the photoproduction of the isolated photon with 
a large $p_T$ at the HERA $ep$ collider is presented. The single resolved 
photon contribution and the QCD corrections 
of order $\alpha_s$ to the Born term are consistently included. The NNLO 
contributions, the box and the double resolved photon subprocesses, are 
sizeable and are taken into account in addition. The importance 
of the isolation cut, as well as the influence of other experimental cuts on 
the $p_T$ and $\eta_{\gamma}$ (the final photon rapidity)
distributions are discussed in detail. The 
investigation of the renormalization scale dependence is performed in order 
to estimate the size of missing higher order QCD corrections.
Results are compared with experimental data and with the prediction of a 
different NLO calculation.}]

\section{Introduction}
The production of the prompt photon with large transverse momentum, $p_T$, 
in the $ep$ collision is considered.
Such reaction is dominated by events with 
almost real photons mediating the $ep$ interaction, 
$Q^2\approx 0$, so in practice we deal with the photoproduction of the 
prompt photon. The other name for such process is 
Deep Inelastic Compton (DIC) scattering (although $Q^2\approx 0$,
the scattering is "deep inelastic" due to a large
transverse momentum of the final photon).
The photon emitted by the electron
may interact with the proton partons directly
or as a resolved one. Analogously, 
the observed final photon
may  arise directly from hard partonic subprocesses 
or from fragmentation processes, 
where a quark or a gluon decays into the photon.

The importance of the DIC process in the $ep$ 
collision for testing the Parton Model and then the Quantum Chromodynamics 
was studied previously by many
authors~\cite{Bjorken:1969ja}$^-~$\cite{Gordon:1994sm}.
Measurements were performed at the HERA ep collider by the ZEUS 
group~\cite{Breitweg:1997pa}$^-$\cite{Breitweg:2000su},
also the H1 Collaboration has presented preliminary
results~\cite{h1}. In these experiments only events with isolated photons
were included in the analysis, 
i.e. with a restriction imposed on the hadronic energy 
detected close to the photon. 
The corresponding cross sections for the photoproduction 
of an isolated photon and of an isolated photon plus jet were calculated
in QCD in next-to-leading order 
(NLO)~\cite{Gordon:1995km}$^-$\cite{Krawczyk:1998it}.
There also exists analogous calculation for the large-$Q^2$ 
$ep$ collision (DIS events)~\cite{Kramer:1998nb}.

In this paper the results of the NLO QCD calculation for the 
DIC process with an isolated photon at the HERA $ep$ collider
are presented. We consider the parton 
distributions in the photon and parton fragmentation into the photon
as quantities of order $\alpha_{em}$.
Our approach differs from the NLO
approach~\cite{Gordon:1995km}$^-$\cite{Gordon:1998yt} 
by set of subprocesses included in the analysis. 
The comparison of our predictions with the NLO results obtained by 
L.E. Gordon (LG)~\cite{Gordon:1998yt} is presented for cross sections
with kinematical cuts as in the ZEUS Collaboration 
measurements~\cite{Breitweg:2000su}.

The present analysis is the final, much extended and improved
version of the previous one~\cite{Krawczyk:1998it}.
We show results for non-isolated final photon, and we study the
influence of the isolation cut on the production rate of the photon. 
The role of other specific cuts applied by the ZEUS Collaboration
are discussed and the comparison with data~\cite{Breitweg:2000su} is made.
We emphasize the importance of the box diagram $\gamma g\rightarrow \gamma g$,
being the higher order process, in description of the data.

We study the renormalization scale dependence of the cross section in order to
estimate the size of missing higher order (NNLO or higher) QCD corrections.
The NLO results for the photoproduction of the isolated $\gamma$
are compared to the leading logarithm (LL) ones, and in addition
the LL predictions for isolated $\gamma + jet$ final state are presented.

In the recent ZEUS analysis of 
the prompt photon plus jet production~\cite{:2001aq} the intrinsic transverse 
momentum of partons in the proton was included in Monte Carlo simulations
to improve agreement between data and predictions. 
This momentum is not included in our calculations.

We start with discussion on the choice of relevant diagrams defining
our NLO approach to the DIC process (sec.~\ref{sec:dic}).
The isolation of the photon is described in sec.~\ref{sec:iso}, and
the equivalent photon approximation in sec.~\ref{sec:epa}.
In secs.~\ref{results} and \ref{th} the results of numerical calculations
are presented and compared with data~\cite{Breitweg:2000su} and other
NLO predictions~\cite{Gordon:1998yt}. In sec.~\ref{ll} we show LL predictions
for the photon plus jet production. Finally, sec.~\ref{sec:sum} 
summarizes our results.

\section{The NLO calculation for $\gamma p\rightarrow\gamma X$ Deep Inelastic
Compton scattering}\label{sec:dic}
\subsection{General discussion} 
\label{sec:disc}

We start  by describing processes which are (should be?) included in the 
NLO QCD calculations of the cross section for  the DIC process, 
\be
\gamma p \rightarrow \gamma X,
\label{eq:1}
\ee
where the final photon is produced with large transverse momentum, 
$p_T\gg\Lambda_{QCD}$.
Although we will consider the process (\ref{eq:1}), the problem which we
touch upon is more general - it is related to different approaches to   
NLO calculations of cross sections for hadronic processes involving resolved 
photons, see~\cite{Krawczyk:1990nq,Krawczyk:1998it} 
and for more detailed discussion~\cite{chyla}.

\begin{figure}
\center
\vskip 0.5cm
\psfig{figure=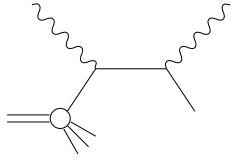,height=1.5cm}
\caption{The Born contribution.}
\label{fig:born}
\end{figure}

The Born level contribution to the cross section for process (\ref{eq:1}),
i.e. the lowest order in the strong coupling $\alpha_s$ term, arises from the 
Compton process on the quark (fig.~\ref{fig:born}):
\be
\gamma q\rightarrow \gamma q.
\label{eq:2}
\ee
It gives the [$\alpha_{em}^2$] order contributions to the partonic cross 
section~\footnote{We 
denote the order of partonic subprocesses using square brackets.}. 
At the same $\alpha_{em}^2$ order it contributes to the hadronic 
cross section for the process $\gamma p\rightarrow \gamma X$.

\begin{figure}
\center
\vskip 1cm
\psfig{figure=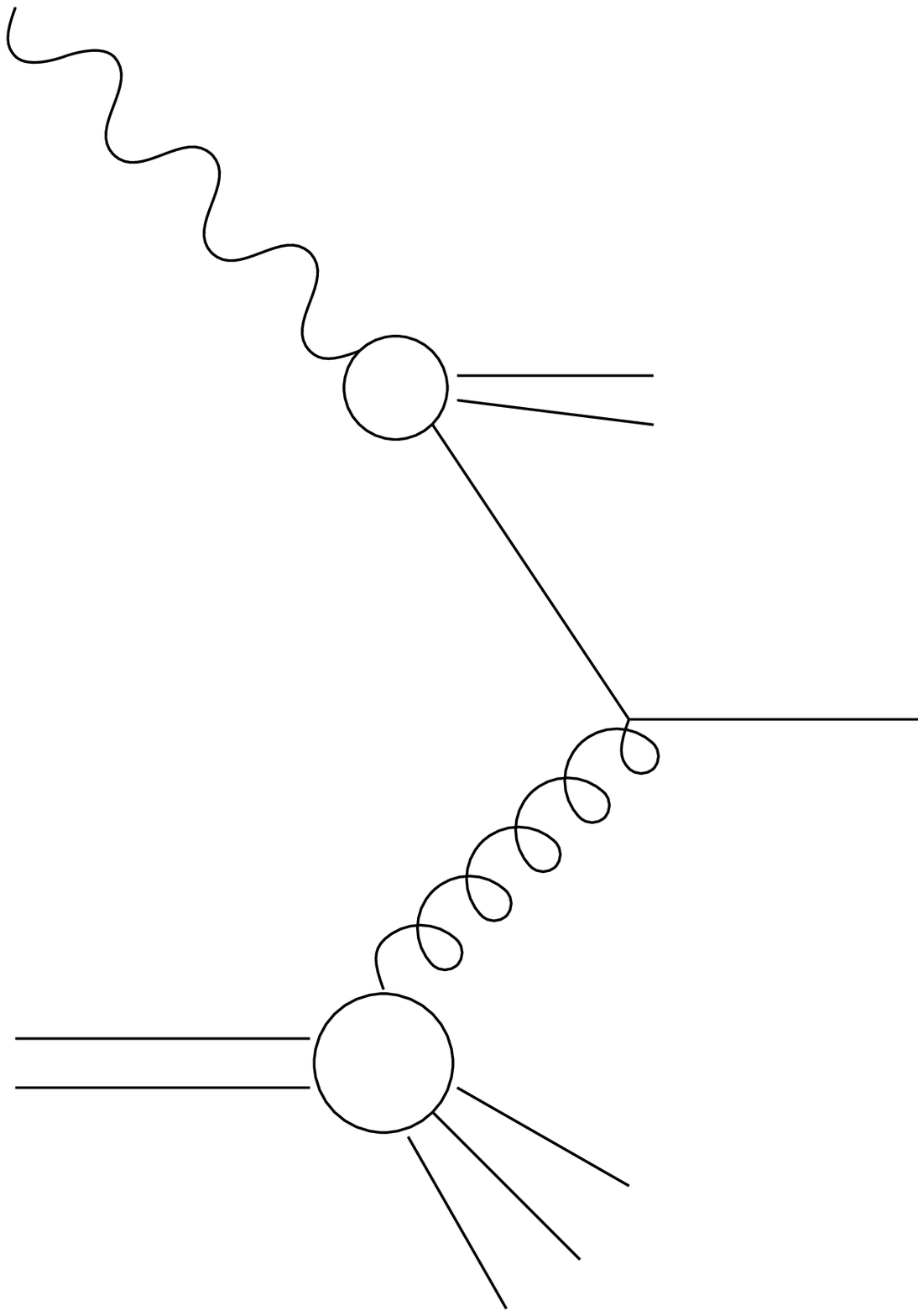,height=1.5cm}
\vskip -1.5cm
\psfig{figure=frag.ps,height=1.5cm}
\caption{Examples of single resolved processes:
a) the resolved initial photon, b) the resolved final photon.}
\label{fig:1res}
\end{figure}

\begin{figure}
\center
\vskip 1cm
\psfig{figure=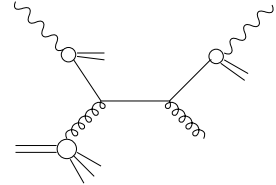,height=1.5cm}
\caption{An example of a double resolved photon process.}
\label{fig:2res}
\end{figure}

The Parton Model (PM) prediction for the DIC process (\ref{eq:1}),
which applies for $x_T=2p_T/\sqrt{S}\sim{\cal O}(1)$,
relies solely on the Born contribution 
(\ref{eq:2})~\cite{Bjorken:1969ja}, namely:
\be
d\sigma^{\gamma p\rightarrow\gamma X}
=\sum_q\int dx_p q_p(x_p){d\hat {\sigma}^{\gamma q\rightarrow \gamma q}},
\label{eq:3}
\ee
where $q_p$ is the quark density in the proton and 
$d\sigma^{\gamma p\rightarrow\gamma X}$ 
($d\hat\sigma^{\gamma q\rightarrow \gamma q}$)
stands for the hadronic (partonic) cross section.
In the QCD improved PM the cross section is
given by (\ref{eq:3}), however with scale dependent quark densities. 
For semihard processes, where $x_T \ll 1$, the prediction
based on the process (\ref{eq:2}) only is not a sufficient approximation, 
and one
should also consider the contributions corresponding to the collinear showers, 
involving hadronic-like interactions of the photon(s). 
There are two classes of such contributions:
{\sl single resolved} with resolved initial $or$ 
final photon, and {\sl double resolved} with
both the initial $and$ the final photon resolved
(figs. \ref{fig:1res}, \ref{fig:2res}).
They correspond to partonic cross sections of orders 
[$\alpha_{em} \alpha_s$] (single resolved)
and [$\alpha_s^2$] (double resolved).
If one takes into account that partonic densities in the photon and
the parton fragmentation into the photon are of order $\sim\alpha_{em}$, 
then the contributions to the hadronic cross section from these resolved 
photon processes are  $\alpha_{em}^2 \alpha_s$ and 
$\alpha_{em}^2 \alpha_s^2$, respectively. 
Both single and double resolved contributions are included in the standard LL 
QCD analyses of the DIC 
process~\cite{Duke:1982bj,Aurenche:1984hc,Gordon:1994sm}. 

\begin{figure}
\center
\vskip 0.5cm
\psfig{figure=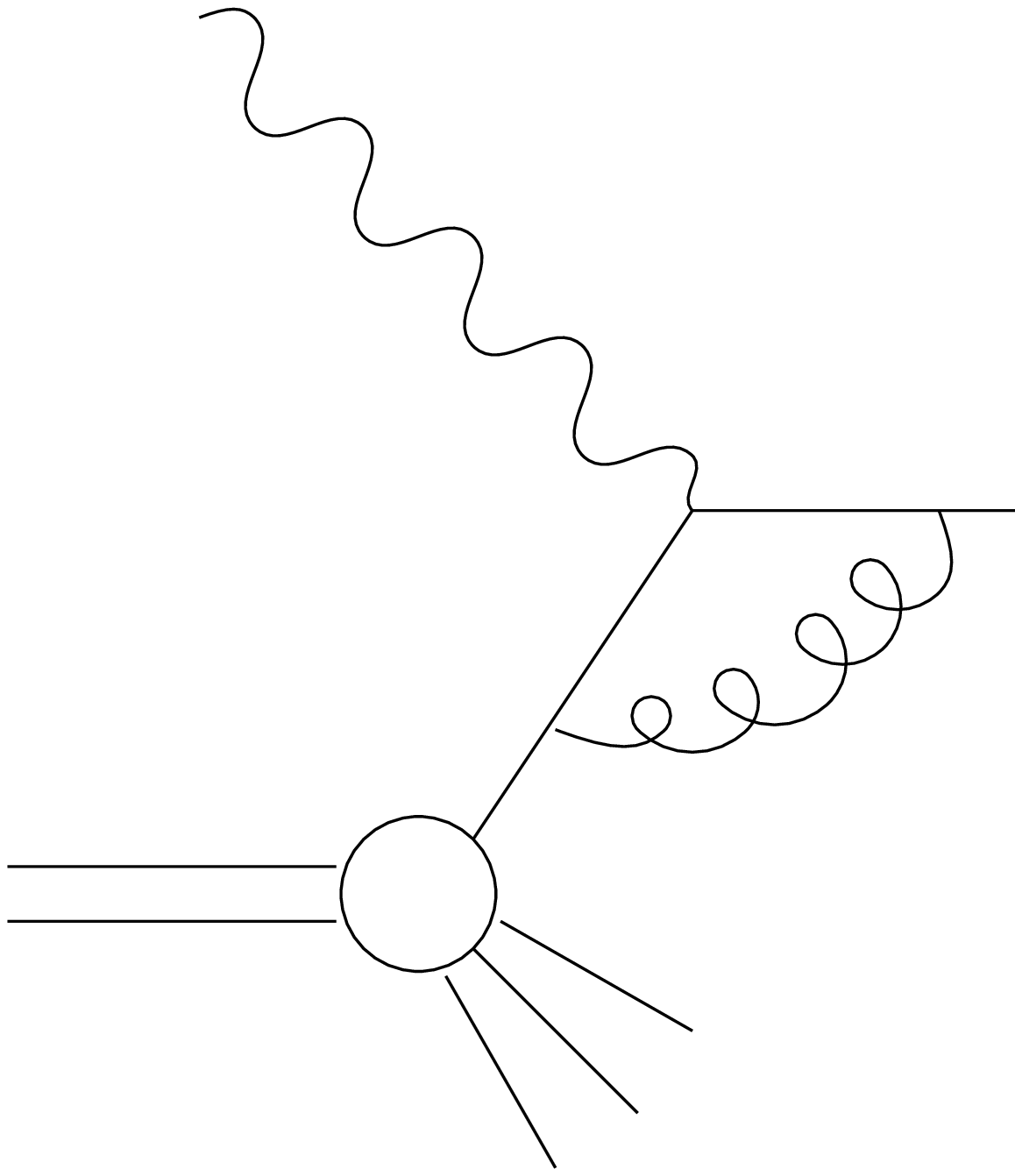,height=1.5cm}
\vskip -1.5cm
\psfig{figure=real.ps,height=1.5cm}
\caption{Examples of the virtual gluon and real gluon $\alpha_s$ corrections
to the Born contribution.}
\label{fig:cor}
\end{figure}

To obtain the NLO QCD predictions for the process (\ref{eq:1})
the $\alpha_s$ corrections to the lowest order process (\ref{eq:2}) 
have to be calculated leading to terms of order 
$\alpha_{em}^2\alpha_s$~\cite{Duke:1982bj,Aurenche:1984hc,mkcorr,jan} 
(fig.~\ref{fig:cor}).
In these $\alpha_{em}^2 \alpha_s$ contributions
there are collinear singularities to be subtracted and 
shifted into corresponding quark densities {\sl or} fragmentation functions.
This way the single resolved photon contribution appears
in the calculation of the $\alpha_s$ corrections to the Born process. 
It is worth noticing that in
the NLO expression for the cross section there are no 
collinear singularities which would lead to the double
resolved photon contributions.
It indicates that taking into account [$\alpha_s^2$] subprocesses,
associated with both the initial and final photons resolved,
goes beyond the accuracy of the NLO calculation.
This will be consistent
within the  NNLO approach, where $\alpha_s^2$ correction to the Born term and
$\alpha_s$ correction to the single resolved terms should be included,
all giving the same  $\alpha_{em}^2\alpha_s^2$ order contribution to the
hadronic cross sections.

The other set of diagrams is considered by some 
authors~\cite{Gordon:1995km}$^-$\cite{Gordon:1998yt} 
in the NLO approach to DIC process 
(\ref{eq:1}), due to their different 
 way of counting the order of parton densities in the  photon 
(and the parton fragmentation into the photon). 
This approach, which we will call
``$1/\alpha_s$'' approach, is motivated by 
large logarithms of $Q^2$ in the $F_2^{\gamma}$ existing already in the 
PM.
By expressing  $\ln (Q^2/\Lambda_{QCD}^2)$ as $\sim{1/\alpha_s}$ 
one treats
the parton densities in photon as proportional to  $\alpha_{em}/\alpha_s$
(see e.g.~\cite{Fontannaz:1982et}$^-$\cite{Aurenche:1984hc,Aurenche:1992sb,Gordon:1994sm,Gordon:1995km}$^-$\cite{Gordon:1998yt}).
By applying this method to the DIC process, we see that
the single resolved photon contribution to the hadronic cross section
for $\gamma p\rightarrow\gamma X$
becomes of the same order as the Born term, namely 
\be
{{\alpha_{em}}\over{\alpha_s}}\otimes [  \alpha_{em} \alpha_s]  
\otimes 1=\alpha_{em}^2.
\ee
The same is also observed for the 
double resolved photon contribution 
\be
{{\alpha_{em}}\over{\alpha_s}}\otimes [ \alpha_s^2]  \otimes 
 {{{\alpha_{em}}}\over{\alpha_s}}=
\alpha_{em}^2.
\ee

We see that by such counting,
 the same $\alpha_{em}^2$ order contributions
to the hadronic cross section are given by the direct Born process,
single and double resolved photon processes
although  they correspond to    quite different final states 
(observe a lack of the remnant of the photon in the direct process).
Moreover, they constitute the lowest order
(in the strong coupling constant) term in the perturbative expansion,
actually the zeroth order, so 
the direct dependence of the cross section
on the strong coupling constant is absent.
If one takes into account that some of these terms  correspond to 
the hard processes involving gluons, the lack of terms proportional
to $\alpha_s$ coupling 
in the cross section seems to be contrary to intuition.

In the ``$1/\alpha_s$'' approach beside  the $\alpha_s$ correction to
the Born cross section, the $\alpha_s$ corrections to the single and to the 
double resolved photon contributions are included in the NLO calculation,
since all of them give terms of the same order,
$\alpha_{em}^2\alpha_s$~\cite{Gordon:1995km}$^-$\cite{Gordon:1998yt}.

To summarize, the first approach starts with one basic, direct subprocess 
as in the PM (eq. \ref{eq:2}), while the second one with three different 
types of subprocesses (as in the standard LL calculation). Obviously, some of 
NNLO terms in the first method belong to the NLO terms in the second one.

In this paper we apply the first type of NLO approach to the DIC process,
however with some important NNLO terms additionally included.
A comparison between our results and the results
based on the other approach~\cite{Gordon:1995km}$^-$\cite{Gordon:1998yt} 
is discussed in sec.~\ref{results5}.

\subsection{The cross section}
\label{sec:cross}

Below we describe our approach to the DIC process,
where the parton densities in the photon and the parton fragmentation 
into the photon are treated as $\sim\alpha_{em}$.

In the NLO QCD calculation of the DIC process we take
into account the following subprocesses:\\
$\bullet$ the Born contribution (\ref{eq:2}) (fig.~\ref{fig:born});\\
$\bullet$ the finite
$\alpha_s$ corrections to the Born diagram (so called K-term) from
virtual gluon exchange, real gluon emission (fig.~\ref{fig:cor}), and 
the process $\gamma g \rightarrow q \bar q \gamma$; \\
$\bullet$ two types of single resolved photon contributions, 
with resolved initial and final photons
(fig.~\ref{fig:1res}).

\begin{figure}
\center
\vskip 1.5cm
\psfig{figure=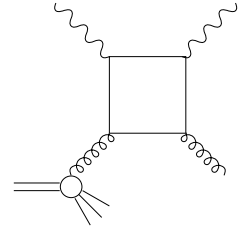,height=1.5cm}
\caption{The box diagram.}
\label{fig:box}
\end{figure}

Besides the above full NLO set, we will include two terms of order 
$\alpha_{em}^2 \alpha_s^2$ (formally  from the NNLO set): the double 
resolved contributions (fig.~\ref{fig:2res}) and the direct diagram (box)
$\gamma g \rightarrow \gamma g$~\cite{Combridge:1980sx} (fig.~\ref{fig:box}), 
since they were found to be 
large~\cite{Fontannaz:1982et}$^-$\cite{Aurenche:1992sb}. 

The cross section for the
$\gamma p\rightarrow\gamma X$ scattering has the following form:
\bea
E_{\gamma}{d^3\sigma^{\gamma p\rightarrow\gamma X}\over
d^3p_{\gamma}} = \sum_{b}\int dx f_{b/p}(x,\bar{Q}^2) 
{\alpha_s(\bar Q^2)\over 2\pi^2 \hat{s}} K_b + \nonumber\\
+\sum_{abc}\int {dz\over z^2}\int dx_{\gamma}\int dx 
f_{a/\gamma}(x_{\gamma},\bar{Q}^2) 
f_{b/p}(x,\bar{Q}^2)
\cdot \nonumber \\ \cdot
D_{\gamma /c}(z,\bar{Q}^2)
E_{\gamma}{d^3\sigma^{ab\rightarrow cd}\over d^3p_{\gamma}}
\eea
The first term is the K-term describing the finite $\alpha_s$ corrections
to the Born process, and the second one stands for the sum over all other
contributions (including the Born contribution). 
The $f_{a/\gamma}$ ($f_{b/p}$) is a $a$ ($b$)-parton 
distribution in the photon (proton) while the $D_{\gamma /c}$ is a $c$-parton 
fragmentation function. For the direct initial (final) photon, 
where $a=\gamma$ ($c=\gamma$), we take $f_{a/\gamma}=\delta (x_{\gamma}-1)$
($D_{\gamma /c}=\delta (z-1)$) 
(the Born contribution is obtained for $a=\gamma$, $b=q$ and $c=\gamma$).
The variables $x_{\gamma}$, $x$ and $z$ 
stand for the fraction of the initial photon,
proton, and $c$-parton momenta taken by the $a$-parton, $b$-parton,
and the final photon, respectively.
The renormalization scale is assumed equal to the factorization 
scale and is denoted as $\bar Q$.

\section{The isolation}
\label{sec:iso}

In order to observe photons originating from a hard subprocess
one should reduce backgrounds, mainly from $\pi^0$'s and $\gamma$'s
radiated from final state hadrons. To achieve this, isolation cuts on the
observed photon are introduced in experimental analyses. The isolation 
cuts are defined by demanding that the sum of 
hadronic energy within a cone of radius $R$ around the final 
photon, where
the radius $R$ is defined in the rapidity and azimuthal
angle space (see eq.~\ref{eq:r} in Appendix),
should be smaller than the final photon energy multiplied
by a small parameter $\epsilon$:
\bea
\sum_{hadrons}E_h<\epsilon E_{\gamma}.
\eea 

The simplest way to calculate the differential cross section
for an isolated photon, $d\sigma_{isol}$, is to calculate
the difference of a non-isolated  differential cross section,
$d\sigma_{non-isol}$, and a subtraction term,
$d\sigma_{sub}$~\cite{Berger:1990es}$^-$\cite{Gluck:1994iz,Gordon:1995km}:
\bea
d\sigma_{isol}=d\sigma_{non-isol}-d\sigma_{sub}.
\eea
The subtraction term corresponds to cuts opposite to the isolation cuts, i.e.
hadrons with the total energy higher than the photon energy multiplied by 
$\epsilon$ should appear within a cone of radius $R$ around the final photon.

The isolation cuts are imposed only when calculating the K-term, and 
in contributions involving fragmentation function (resolved final photon). 
Other contributions arise from $2\rightarrow 2$ subprocesses with direct 
final photon that is isolated by definition.

In the analysis we apply the subtraction method with the subtraction 
term calculated in an approximate way, 
see~\cite{Gordon:1994ut,Gordon:1995km} for details. 
The approximation bases on the assumption that  
an angle $\delta$ between the final photon and 
a parton inside the cone of radius $R$ is small.
It allows to simplify considerably the calculations
and leads to compact analytical expressions for all relevant
matrix elements involved in $d\sigma_{sub}$. 
Note that in this approximation the angle $\delta$ is simply proportional 
to the radius $R$: $\delta=R/cosh(\eta_{\gamma})$.
The above small-$\delta$ approximation is used only in calculation of the 
$K$-term in the subtraction cross section $d\sigma_{sub}$,
for the results see Appendix. Other 
contributions to $d\sigma_{sub}$, as well as $d\sigma_{non-isol}$
and all LL expressions, are obtained in an exact way.

It is worth mentioning that there is an ongoing discussion whether
the conventional factorization breaks down, and whether the cross section
is an infrared safe quantity for isolated 
photon photoproduction in $e^+e^-$ collisions (also for hadron-hadron 
reactions)~\cite{Berger:1996cc,Aurenche:1997ng}.
In principle these questions could as well  
occur for the photoproduction of isolated photons in $ep$ collisions.
However we do not deal with this problem
because it arises from $2\rightarrow 3$ subprocesses in which
a final quark fragmentates into a photon. 
We checked this explicitly and found that all singularities in
$d\sigma_{sub}$ are canceled or factorized, as in 
$d\sigma_{non-isol}$. Therefore the considered by us cross section 
$d\sigma_{isol}$ is well defined (see 
also~\cite{Gordon:1995km,Gordon:1998yt,Kramer:1998nb}).

\section{The equivalent photon approximation}\label{sec:epa}

We consider the production of photons with large transverse momentum,
$p_T\gg\Lambda_{QCD}$, in the $ep$ scattering, $ep\rightarrow e\gamma X$,
at the HERA collider. This reaction is dominated by photoproduction
events, i.e. the electron is scattered at a small angle and the
mediating photon is almost real, $Q^2\approx 0$. The cross section for 
such processes can be calculated using the equivalent photon
(Williams-Weizs\"{a}cker) approximation~\cite{vonWeizsacker:1934sx} 
which relates the differential cross section for $ep$ collision
to the differential cross section for $\gamma p$ 
collision. For the DIC scattering the approximation has the following form:
\bea
d{\sigma}^{ep\rightarrow e\gamma X}=\int G_{\gamma/e}(y) 
d{\sigma}^{\gamma p\rightarrow\gamma X} dy ,
\eea
where $y$ is (in the laboratory frame) a fraction of the initial electron 
energy taken by the photon. 

We apply the equivalent photon approximation and
take the (real) photon distribution in the electron in the 
form~\cite{Budnev:1974de}:
\bea
G_{\gamma/e}(y)={\alpha_{em}\over 2\pi} \{ {1+(1-y)^2\over y}
\ln [ {Q^2_{max}(1-y)\over m_e^2 y^2}] \nonumber \\
- ~ {2\over y}(1-y-{m_e^2y^2\over Q^2_{max}}) \},
\eea
with $m_e$ being the electron mass. In the numerical calculations
we assume $Q^2_{max}$ as 1 GeV$^2$ what is a typical value for the 
recent photoproduction measurements at the HERA collider.

\section{The results and comparison with data}\label{results}

The results for the non-isolated and isolated photon cross sections are 
obtained in NLO accuracy with additional NNLO
terms, as discussed in sec.~\ref{sec:cross}. 
We take the HERA collider energies: $E_e$=27.5 GeV and 
$E_p$=820 GeV~\cite{Breitweg:2000su},
and we consider the $p_T$ range of the final photon between 
5 and 20 GeV ($x_T$ from 0.03 to 0.13).
The calculations are performed in ${ \overline {\rm MS}}$ scheme with
a hard (renormalization, factorization) scale $\bar{Q}$ equal $p_T$.
Also $\bar{Q}=p_T/2$ and $2p_T$ are used to study the dependence
of the results on the choice of $\bar{Q}$.
We neglect the quark masses and assume
the number of active flavors 
to be $N_f$=4 (and for comparison also $N_f$=3 and 5).
The two-loop coupling constant $\alpha_s$ is used in the form
\be
\alpha_s(\bar Q^2)={{4 \pi}\over {\beta_0 \ln(\bar{Q}^2/\Lambda_{QCD}^2) }}
[1-{{2\beta_1}\over{\beta_0^2}} {{\ln[\ln(\bar{Q}^2/\Lambda_{QCD}^2)]}
\over{\ln(\bar{Q}^2/\Lambda_{QCD}^2)}}]
\ee
($\beta_0=11-2/3 N_f$ and $\beta_1=51-19/3 N_f$),
with $\Lambda_{QCD}$=0.365, 0.320 and 0.220 GeV for $N_f$=3, 4 and 5,
respectively, as fitted by us 
to the experimental value of $\alpha_s(M_Z) = 0.1177$~\cite{Biebel:1999zt}.
The $\Lambda_{QCD}^{LL}$ = 0.120 GeV
for $N_f$=4 was taken in one-loop $\alpha_s$ 
when calculating the cross section in LL accuracy.

We use the GRV parametrizations of the proton
structure function (NLO and LO)~\cite{Gluck:1995uf}, 
the photon structure function (NLO and LO)~\cite{Gluck:1992ee},
and the fragmentation function (NLO)~\cite{Gluck:1993zx}. For comparison
other parametrizations are also 
used: DO~\cite{Duke:1982bj}, ACFGP~\cite{Aurenche:1992sb}, 
CTEQ~\cite{Lai:1997mg}, MRST~\cite{Martin:1998sq} and GS~\cite{Gordon:1997pm}.

As the reference we take the GRV NLO set of parton 
distributions~\cite{Gluck:1995uf}$^-$\cite{Gluck:1993zx},
$N_f = 4$, $\Lambda_{QCD} = 320$ GeV and $\bar{Q}=p_T$.

\subsection{Non-isolated versus isolated photon cross section}
\label{results1}

\begin{figure*}
\center
\vskip 0.5cm
\psfig{figure=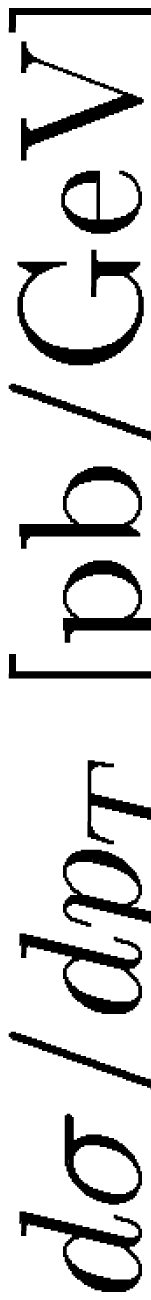,height=7cm}
\caption{The final photon $p_T$-dependence of the
cross section $d\sigma /dp_T$ for non-isolated $\gamma$
photoproduction (solid line).
The Born contribution is shown separately 
(dash-dotted line).}
\label{fig:dptincl}
\end{figure*}

The $p_T$ distribution for the produced final photon without any cut is 
presented in fig.~\ref{fig:dptincl} where the NLO results and separately 
the Born term (with NLO parton densities) are shown.
The cross section decreases by three orders of magnitude when $p_T$
increases from 4 GeV to 20 GeV, and obviously the most important 
contribution is coming from the lowest $p_T$ region.
The subprocesses other than the Born one give all together
contribution almost two times
larger than the cross section for the Born subprocess alone.

The importance of particular contributions to the non-isolated cross section 
integrated over 5 GeV$<p_T<$ 10 GeV is illustrated in Table~1 (the first line).
The total NLO cross section is equal to 226 pb, with  
individual contributions  equal to:
$Born=36.3\% $, $single$ $resolved=35.1\% $,
$double$ $resolved=18.7\%$, $box=6.2\% $, $K$-$term$=3.9\%.
We see that the single resolved photon processes
give a contribution comparable to the Born term. Also the double resolved
 photon processes are  important. It is worth noticing that the overall 
double resolved photon cross section is build from   many, relatively small, 
individual terms. The direct box diagram ($\gamma g\rightarrow\gamma g$) 
gives 17\% of the Born ($\gamma q\rightarrow\gamma q$) contribution.
Relatively large box contribution (although being $[\alpha_s^2]$)
is such partially due to large gluonic content of the proton at small $x_p$.

\begin{figure*}
\center
\vskip 0.5cm
\psfig{figure=R.ps,height=7cm}
\vskip -7cm
\psfig{figure=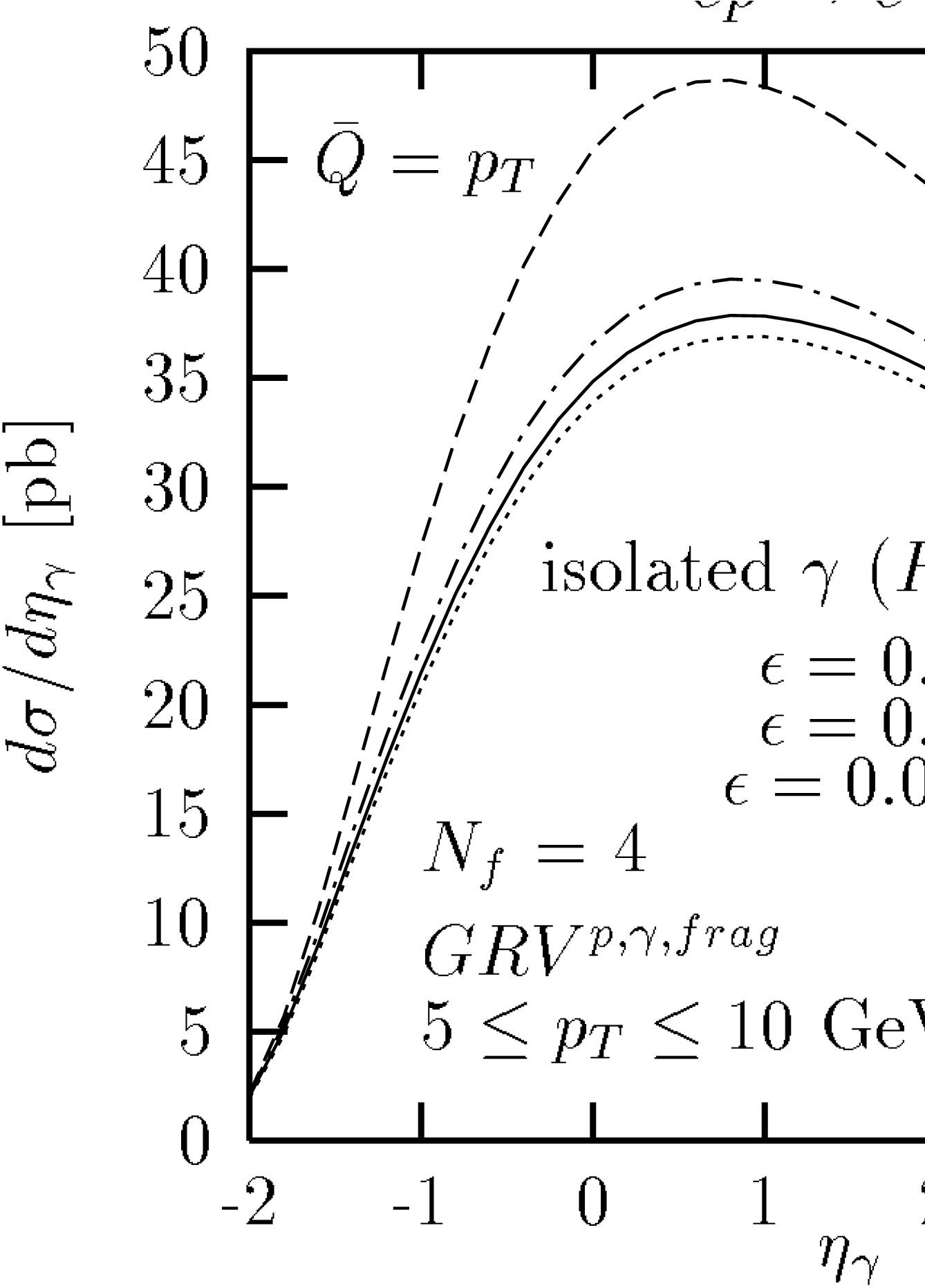,height=7cm}
\caption{The differential cross section $d\sigma/d\eta_{\gamma}$ as a function
of the photon rapidity $\eta_{\gamma}$.
a) The results for non-isolated photon (dashed line) and isolated photon
with $R = 1$ and $\epsilon =$ 0.05 (dotted line), 0.1 (solid line)
and 0.2 (dot-dashed line).
b) The results for non-isolated photon (dashed line) and isolated photon
with $\epsilon = 0.1$ and $R =$ 0.1 (dot-dashed line), 0.5 (dotted line)
and 1 (solid line).
}
\label{fig:Reps}
\end{figure*}

Next, in fig.~\ref{fig:Reps} we compare the differential cross section 
$d\sigma /d\eta_{\gamma}$ for the non-isolated photon with 
corresponding predictions for the isolated photon using various values of 
the isolated cone variables ($\epsilon$, $R$)~\footnote{The positive
rapidity is pointed in the proton direction.}.
The isolation cut suppresses the cross section by above 10\% in the whole 
rapidity range. For $\epsilon$=0.1 and $R=1$
the suppression is 17-23\% at rapidities $-1.5<\eta_{\gamma}\le 4$.
This large effect is not too sensitive to the value of $\epsilon$:
changing the value by a factor of 2 from $\epsilon =0.1$ to
$\epsilon =0.2$ or to $\epsilon =0.05$ varies the results for isolated photon
by about $4\%$ (fig.~\ref{fig:Reps}a).
The dependence on $R$ is stronger but also not very large: when changing
$R$ value by a factor of 2 from 1 to 0.5 the results increase by about $7\%$
(fig.~\ref{fig:Reps}b).

The suppression due to the isolation imposed on the photon is presented
in Table~1 (the second line) for individual contributions and for the total 
cross section. As expected, the cross section for fragmentation processes 
(i.e. with resolved final photon) is strongly suppressed: 
after isolation it is lowered by a factor of 5. 
At the same time the QCD corrections to the Born diagram increase
significantly, i.e. the contribution to the subtraction
cross section, $d\sigma_{sub}$, due to this corrections is negative.
The isolation restrictions do not modify contributions of other subprocesses 
since they involve photons isolated by definition.
The subtraction cross section, being a sum of negative QCD corrections
and fragmentation contributions, is of course positive and the
total cross section for isolated final photon is lower, by $20\%$,
than for the non-isolated one.

In the following we keep $R$=1 and $\epsilon$=0.1,
standard values used in both theoretical and experimental analyses.

\subsection{Other experimental cuts}\label{results1b}

\begin{figure*}
\vskip -0.8cm
\hskip 4.6cm
\psfig{figure=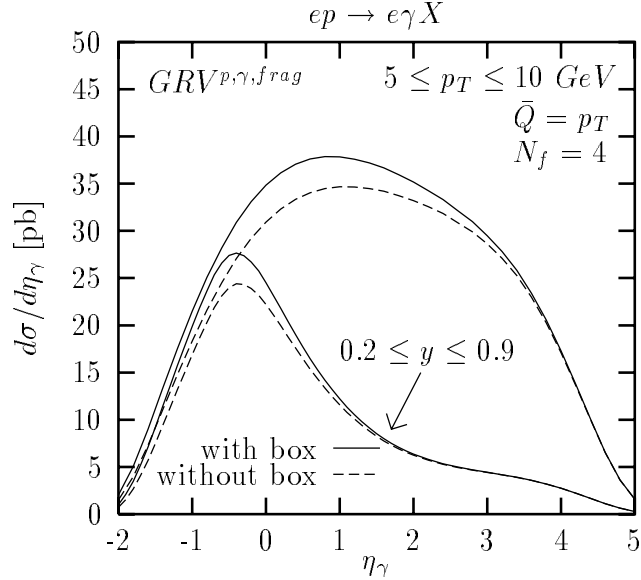,height=7cm}
\caption{The differential cross section $d\sigma/d\eta_{\gamma}$ 
for isolated $\gamma$ ($\epsilon = 0.1$, $R = 1$)
as a function of the photon rapidity $\eta_{\gamma}$ 
with (solid lines) and without (dashed lines) the box contribution. 
The results are obtained with imposed $y$ cut ($0.2\leq y\leq 0.9$)
and without this cut.}
\label{fig:cd}
\end{figure*}

In order to compare the results with data we consider other 
cuts imposed by the ZEUS group on prompt photon events at the HERA 
collider~\cite{Breitweg:2000su}.
The influence of the limited energy range, $0.2\le y\le 0.9$,
is shown in fig.~\ref{fig:cd}.
The cross section is strongly reduced, by 30-85\%, in the positive rapidity 
region. At negative rapidities the change due to the $y$-cut is weaker: 5-10\% 
at $-1.2<\eta_{\gamma}\le -0.4$ and 10-30\% at other negative rapidities.
Separately we show the results obtained without the box 
subprocess (fig.~\ref{fig:cd}). The box diagram 
contributes mainly in the rapidity region between -1 and 3. After 
imposing the $y$-cut its contribution is important in narrower region 
from -1 to 1. The influence of the $y$-cut can be 
read also from Table~1 (the third line).
One sees e.g. that the Born contribution is reduced 3.5 times, 
while other ones are suppressed less, roughly by a factor of 2.

The results obtained for the isolated photon with the $y$-cut and in addition 
with the cut on the final photon rapidity, $-0.7\le\eta^{\gamma}\le 0.9$, 
are presented in the last line of Table~1. The restriction on 
$\eta^{\gamma}$ decreases the contributions of all subprocesses
approximately by a factor of 2 (except for the double resolved contribution
reduced almost 3 times).

\begin{figure*}
\hskip 4.6cm
\psfig{figure=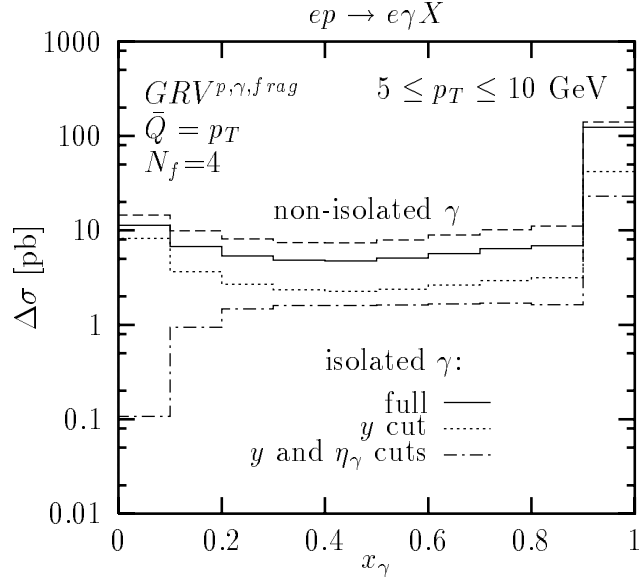,height=7cm}
\caption{The cross section in $x_{\gamma}$ bins of the length 0.1. 
The results for non-isolated $\gamma$ integrated over the whole range 
of $y$ and $\eta_{\gamma}$ are shown with the dashed line. 
The solid line represents results integrated over the whole 
range of $y$ and $\eta_{\gamma}$ for isolated $\gamma$ 
with $\epsilon = 0.1$ and $R = 1$.
Results with additional cuts in the
isolated $\gamma$ cross section are shown with: dotted line
($0.2\leq y\leq 0.9$) and dot-dashed line ($0.2\leq y\leq 0.9$,
$-0.7\leq\eta_{\gamma}\leq 0.9$).}
\label{fig:dxp}
\end{figure*}

The role of various experimental cuts is illustrated
also in fig.~\ref{fig:dxp}, this time for the $x_{\gamma}$ distribution.
In particular we see that the isolation and the energy cut reduce considerably
the contributions from large and medium $x_{\gamma}$, while the contributions
from $x_{\gamma}$ below 0.1
are reduced less. On the other hand, the small $x_{\gamma}$
contributions are strongly, by two orders of magnitude, 
diminished by the photon rapidity cut.
This shows that measurements at the central $\eta^{\gamma}$ region 
($-0.7\le\eta^{\gamma}\le 0.9$)
are not too sensitive to the small $x_{\gamma}$ values in the photon.

When calculating
the QCD corrections to the Born process in the subtraction 
term $d\sigma_{sub}$ we used the small-$\delta$ approximation 
described in sec.~\ref{sec:iso}.
Because these corrections give less than 10\% 
of the cross section for the isolated 
photon production with various cuts (see the third column in Table~1), 
we expect that the error resulting from using the approximations is small, 
though we use in fact not a small value of $\delta$ 
($\delta=R/cosh \eta_{\gamma}$, $R$ = 1).~\footnote{Calculations 
for the prompt photon production in $ep$~\cite{Gordon:1998yt} and
hadron-hadron~\cite{Gordon:1997kc} collisions were performed
using space slicing method without the small $\delta$ assumption.
Comparison of such results with predictions obtained in
(discussed here) approximated way showed
that the small $\delta$ approximation is an accurate
analytic technique for including isolation effects in NLO calculations
(also for $R$ = 1)~\cite{Gordon:1998yt}.}

\subsection{The comparison with data}\label{results2}

Two types of the final state were measured in the ZEUS experiment:
1) an isolated photon with $-0.7\le\eta^{\gamma}\le 0.9$ 
and $5\le p_T\le 10$ GeV; 
2) an isolated photon plus jet with the photon 
rapidity and transverse momentum as above, the jet rapidity in the range
$-1.5\le\eta^{jet}\le 1.8$, and the jet transverse momentum 
$p_T^{jet}\ge 5$ GeV.

\begin{figure*}
\vskip 0.5cm
\psfig{figure=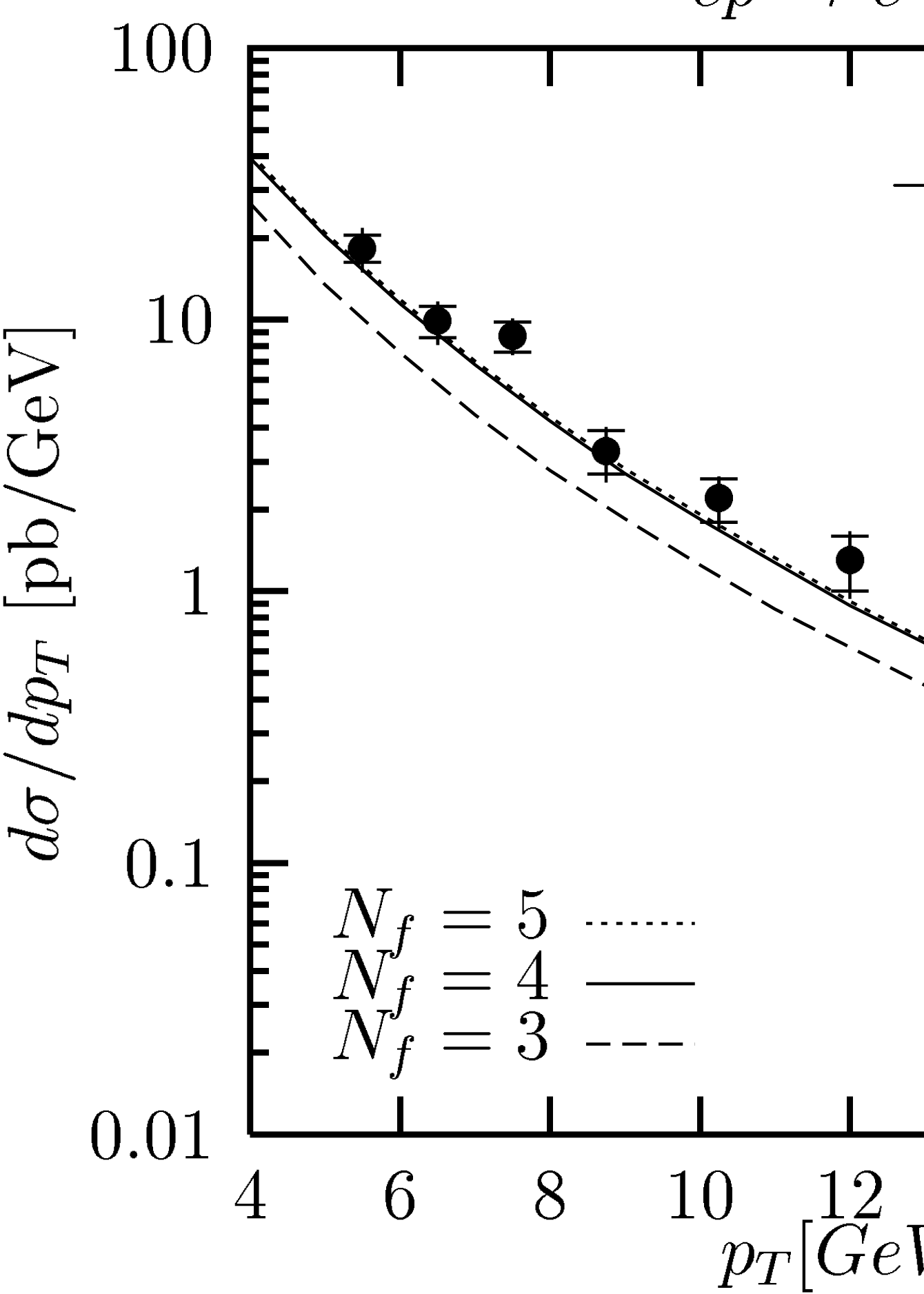,height=7cm}
\vskip -7cm
\psfig{figure=dnNf.ps,height=7cm}
\caption{
The results for isolated $\gamma$
with various numbers of active massless flavors:
$N_f$ = 3 (dashed lines), 4 (solid lines) and 5 (dotted lines),
compared to the ZEUS data~\protect\cite{Breitweg:2000su}.
a) The differential cross section $d\sigma/dp_T$ as a function of the
photon transverse momentum.
b) The differential cross section $d\sigma/d\eta_{\gamma}$
as a function of the photon rapidity $\eta_{\gamma}$;
the result without the box contribution
is also shown for $N_f$ = 4 (dot-dashed line).} 
\label{fig:dnNf}
\end{figure*}

We compare our NLO predictions with the ZEUS 
data from the first type of measurements~\cite{Breitweg:2000su}.
In fig.~\ref{fig:dnNf}a the comparison is made for
the transverse momentum distribution for various $N_f$.
Although the predictions tend to lie slightly below the data a satisfactory 
agreement is obtained for $N_f=4$.
Note a large difference between the results for $N_f$=4 and 3
due to the fourth power of electric charge characterizing  processes 
with two photons.
We observe a very  small contribution from the bottom quark (for $N_f$=5). 
The predictions are obtained in massless quark scheme and may overestimate 
the production rate.

Similar comparison of the NLO results with the data, now for the rapidity 
distribution, is shown in fig.~\ref{fig:dnNf}b. A good description of the 
data is obtained for $N_f$=4 and $N_f$=5 in the rapidity region 
$0.1\le\eta_{\gamma}\le 0.9$. For $-0.7\le\eta_{\gamma}\le 0.1$ our 
predictions lie mostly below the experimental points. 
This disagreement between predicted and measured cross sections is observed 
also for other theoretical calculations (LG) and for Monte Carlo 
simulations~\cite{Breitweg:2000su}~\footnote{The ZEUS Collaboration has 
presented recently an analysis~\cite{:2001aq} in which an intrinsic transverse 
momentum of partons in the proton, $k_T$, was introduced in the PYTHIA 6.1 
generator in order to improve agreement between the data and Monte Carlo 
predictions for an isolated photon plus jet photoproduction. The data, selected
with $x_{\gamma}>0.9$, are consistent with the predictions for 
$<k_T> = 1.69 \pm 0.18 ^{+0.18}_{-0.20}$ GeV.}.
In fig.~\ref{fig:dnNf}b we present separately an effect due to the box 
subprocess (for $N_f=4$). It is clear that the box term enhances considerably 
the cross section in the measured rapidity region. Its contribution to the 
integrated cross section is equal to 9.6\%. The double resolved photon 
contribution is also sizeable, although roughly two times smaller than the 
box one, see Table~1 (fourth line).
Both these $[\alpha_s^2]$ contributions improve description of the data.

\begin{figure*}
\vskip 0.5cm
\psfig{figure=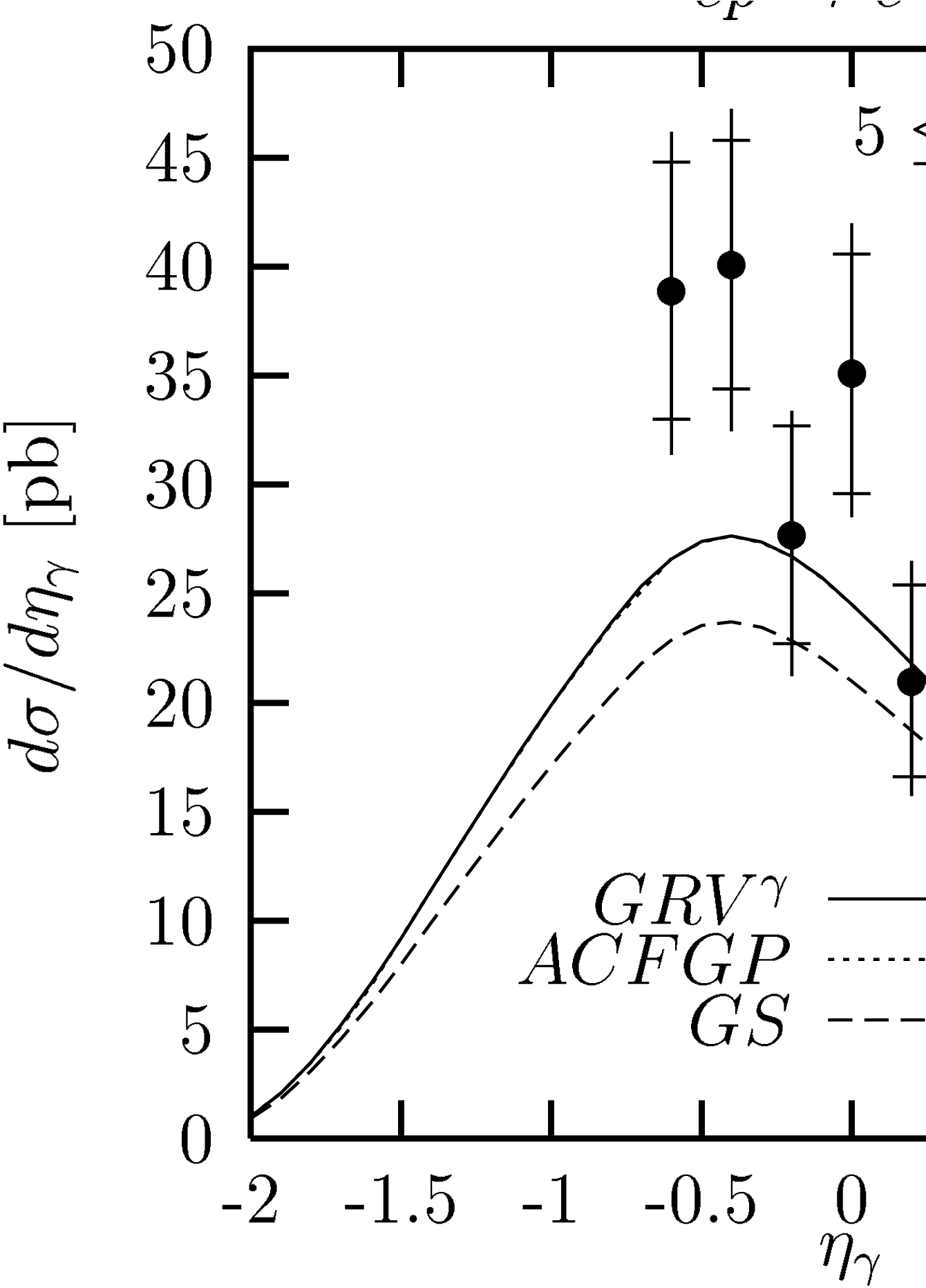,height=7cm}
\vskip -7cm
\psfig{figure=dnpar2.ps,height=7cm}
\caption{The differential cross section $d\sigma/d\eta_{\gamma}$
for isolated $\gamma$
as a function of the photon rapidity $\eta_{\gamma}$ compared to
the ZEUS data~\protect\cite{Breitweg:2000su}.
Three different NLO photon parton distributions are used: 
ACFGP~\protect\cite{Aurenche:1992sb} 
(dotted line), GRV~\protect\cite{Gluck:1992ee} (solid line) and 
GS~\protect\cite{Gordon:1997pm} (dashed line).
The GRV NLO parton distributions in the proton~\protect\cite{Gluck:1995uf}
and parton fragmentation into photon~\protect\cite{Gluck:1993zx} are used.
a) $\bar{Q} = p_T$. b) $\bar{Q} = 2 p_T$.}
\label{fig:dnpar}
\end{figure*}

The predictions obtained using three different NLO parton densities in the 
photon (ACFGP~\cite{Aurenche:1992sb}, GRV~\cite{Gluck:1992ee} and 
GS~\cite{Gordon:1997pm}) are presented for $N_f=4$
in fig.~\ref{fig:dnpar}a ($\bar{Q} = p_T$) and in fig.~\ref{fig:dnpar}b 
($\bar{Q} = 2 p_T$) together with the ZEUS data~\cite{Breitweg:2000su}.
The results based on ACFGP and GRV parametrizations differ by less than 4\%
at rapidities $\eta_{\gamma} < 1$ (at higher $\eta_{\gamma}$ the difference is 
bigger), and both give good description of the data in the rapidity range
$0.1\le\eta_{\gamma}\le 0.9$ (for $\bar{Q} = p_T$ and $\bar{Q} = 2 p_T$). 
For $-0.7\le\eta_{\gamma}\le 0.1$ none of the predictions is in
agreement with the measured cross section.

For $\bar{Q}$=$p_T$ (fig.~\ref{fig:dnpar}a) the GS distribution leads to 
results considerably below ones obtained using ACFGP and GRV densities,
especially in the rapidity region from roughly -1 to~1.
This difference between the GS and other 
considered herein parton parametrizations is 
mainly due to their different treatment of the 
charm quark in the photon.
In the GS approach the charm quark is absent
for $\bar{Q}^2$ below 50 GeV$^2$.
Since we take $5\leq \bar{Q} = p_T\leq 10$ GeV, and the most important 
contribution to the cross section arises from the lower $p_T$ region 
(see fig.~\ref{fig:dnNf}a), the $\bar{Q}^2$ value 
usually lies below the GS charm quark threshold.
As a consequence, predictions based on GS have strongly suppressed 
the contribution
of subprocesses involving charm from the photon - contrary
to GRV and ACFGP predictions where the charm threshold is at lower $\bar{Q}^2$.

The above explanation of differences between cross sections involving
GS and both GRV and ACFGP parton densities is insufficient for higher 
rapidities, $\eta_{\gamma} > 2$.
Here the differences between the results based on particular 
photon parametrizations are bigger, especially
when comparing predictions obtained using GRV and ACFGP ones (not shown). 
This is due to large differences between used parton densities at 
low $x_{\gamma}$, which is  probed at the high rapidity region.

All the considered parton distributions give similar description of the data
when the scale is changed to $\bar{Q}=2p_T$, see fig.~\ref{fig:dnpar}b. 
Here the calculation corresponds to $\bar{Q}^2$
which is always above 50 GeV$^2$ and 
the charm density in the GS parametrization is non-zero,
as in other parametrizations.

\begin{figure*}
\vskip 0.5cm
\psfig{figure=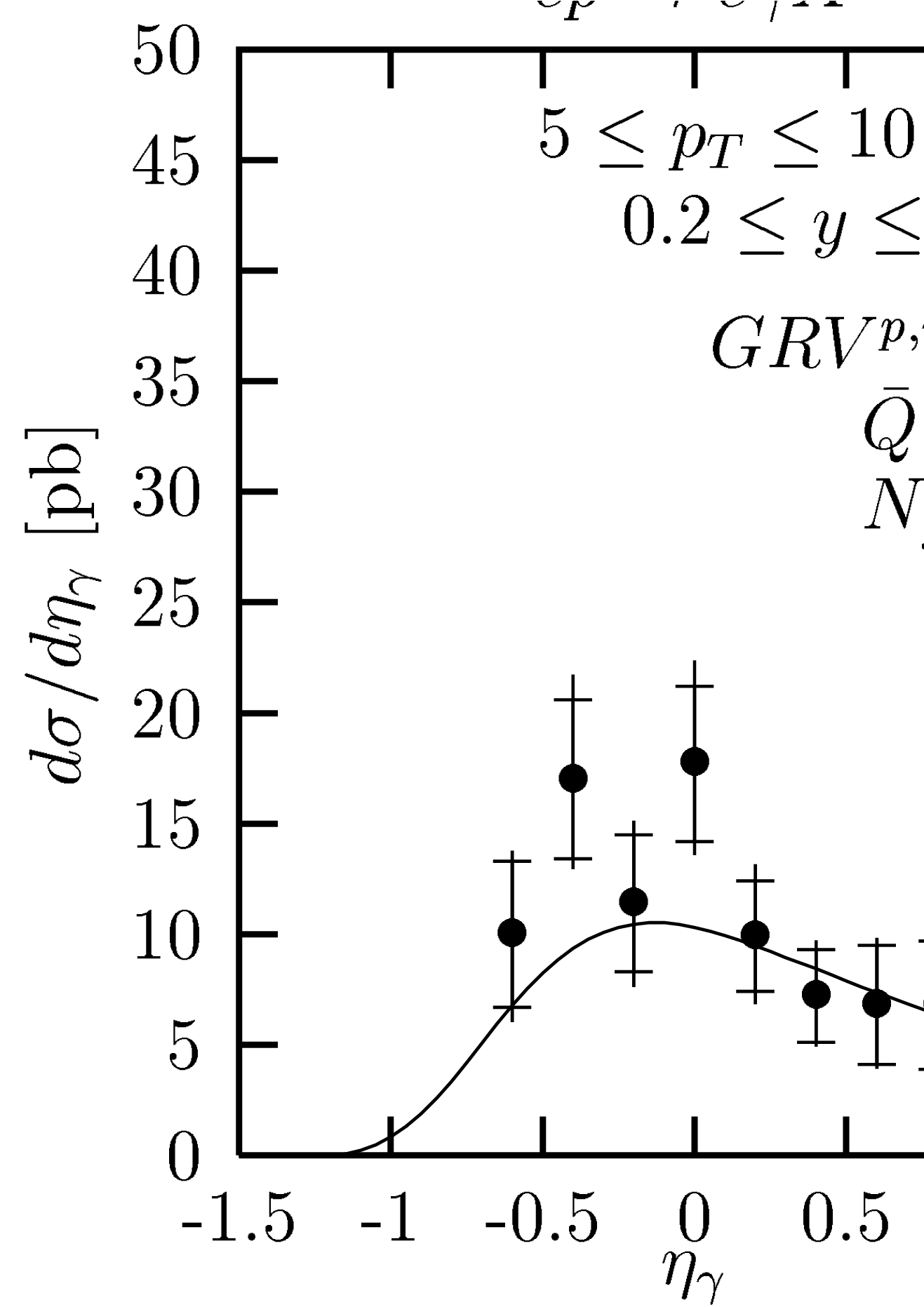,height=7cm}
\vskip -7cm
\psfig{figure=dn_b.ps,height=7cm}
\vskip -7cm
\psfig{figure=dn_c.ps,height=7cm}
\caption{The results for three ranges of $y$:
$0.2<y<0.32$, $0.32<y<0.5$ and $0.5<y<0.9$, compared to the ZEUS 
data~\protect\cite{Breitweg:2000su}.}
\label{fig:bins}
\end{figure*}

In fig.~\ref{fig:bins} our predictions are compared to the ZEUS data
divided into three ranges of $y$. This allows to establish that
the above discussed discrepancy between the data and the predictions for 
$\eta^{\gamma} < 0.1$ is coming mainly from the low $y$ region, $0.2<y<0.32$.
In the high $y$ region, $0.5<y<0.9$, a good agreement is obtained.

We have also studied the dependence of our results on the choice of 
the parton distributions in the proton and parton fragmentation
into the photon (not shown). Cross sections calculated using 
GRV~\cite{Gluck:1995uf}, 
MRST (set ft08a)~\cite{Martin:1998sq} and CTEQ4M~\cite{Lai:1997mg} NLO parton 
parametrizations for the proton vary among one another by 4 to 7\% 
at negative rapidities and less than $4\%$ at positive rapidity values.
Results for the isolated final photon are also not too sensitive to the
fragmentation function. For rapidity ranging from -1 to 4 the cross section
obtained with DO LO~\cite{Duke:1982bj} 
fragmentation function is $2-3.5\%$ lower than
the cross section based on GRV NLO~\cite{Gluck:1993zx} parametrization. 
Only at minimal ($\eta_{\gamma}<-1$)
and maximal ($4<\eta_{\gamma}$) rapidity values
this difference is larger, being at a level of $3.5-8\%$.

\section{The theoretical uncertainties of the results and comparison 
with other NLO predictions}\label{th}

As we already mentioned the predictions are obtained in massless quark 
scheme and may overestimate the production rate.
An improved treatment of the charm quark,
especially in the box contribution which is particularly sensitive
to the change from $N_f=3$ to $N_f=4$, would be needed.
However we do not expect that this improvement would change qualitatively
our results.

We now discuss the theoretical uncertainties of our predictions
related to the perturbative expansion.

\subsection{The dependence on the $\bar{Q}$ scale}\label{results3}

\begin{figure*}
\vskip 0.5cm
\psfig{figure=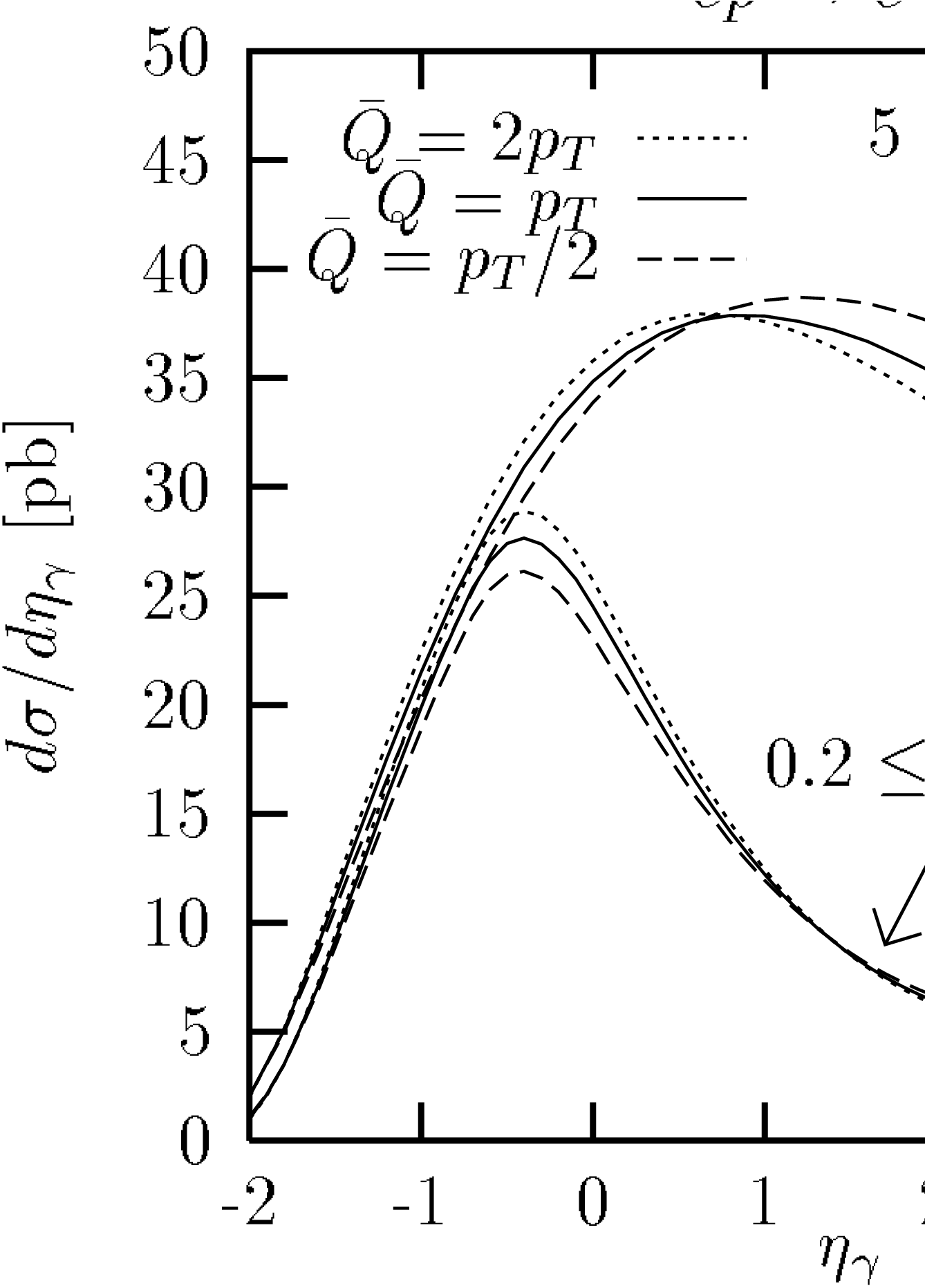,height=7cm}
\vskip -7cm
\psfig{figure=dnQnobox.ps,height=7cm}
\caption{The differential cross section $d\sigma/d\eta_{\gamma}$ 
for isolated $\gamma$ photoproduction
as a function of the photon rapidity $\eta_{\gamma}$ with (a) and 
without (b) the box contribution.
Three different values of the $\bar{Q}$ scale
are assumed: $\bar{Q} = p_T/2$ (dashed lines),
$\bar{Q} = p_T$ (solid lines) and $\bar{Q} = 2 p_T$ (dotted lines).
The results are obtained with imposed $y$ cut ($0.2\leq y\leq 0.9$)
and without this cut.}
\label{fig:dnQ}
\end{figure*}

In order to estimate the contribution due to missing higher order 
terms, the influence of the choice of the $\bar Q$ scale is studied for 
the $\eta_{\gamma}$ distribution.
In fig.~\ref{fig:dnQ}a the results obtained using GRV densities
with and without the $y$-cut are shown.
When changing $\bar Q$ from $p_T$ to 2$p_T$ 
($p_T$/2) the cross section
increases (decreases) at rapidities below $\sim 1$
and decreases (increases) at higher rapidity values. 
Only at high rapidities (where the cross section is small),
$\eta_{\gamma} > 3$, the dependence on the choice 
of the scale is strong, above 10\% 
(up to 20-30\% at $\eta_{\gamma}\approx 5$). 
In the wide kinematical region,
$-2 < \eta_{\gamma} < 2$, the relative differences between results
(with and without the $y$-cut)
for $\bar Q = p_T$ and results for $\bar Q = 2p_T$ or $p_T/2$ 
are small and do not exceed 6\%.
Around the maximum of the cross section at rapidities 
$-1\le\eta_{\gamma}\le 0$ these differences are 4-6\%. 
This small sensitivity of the results to the change of the scale is 
important since it indicates that the contribution from neglected NNLO
and higher order terms is not significant.

Note that individual contributions are strongly dependent
on the choice of $\bar{Q}$, e.g. results for the
single resolved processes vary by
$\pm$10-20\% at rapidities $\eta_{\gamma}\le 1$.
Results are much more stable
only when the sum of resolved processes
and QCD corrections is considered.

In fig.~\ref{fig:dnQ}b we present NLO results for various $\bar{Q}$ 
with and without the $y$-cut, however this time with no box contribution.
At rapidities $\eta_{\gamma}<1$ the uncertainty due to the choice of the 
renormalization scale is about two times higher than for the 
cross section with included box diagram,
so the box contribution ($\sim [\alpha_s^2]$) seems to stabilize 
the NLO prediction. At rapidities $\eta_{\gamma}> 2$ the
relative dependence on the choice of the scale is similar for the
cross section with and without the box term.

\subsection{The comparison of NLO and LL predictions}\label{results4}

In the present calculation we include in LL accuracy the single and 
double resolved photon processes as well as the box diagram in addition
to the Born contribution, see 
also~\cite{Fontannaz:1982et}$^-$\cite{Aurenche:1984hc,Gordon:1994sm} 
(although this is not fully consistent with 
the discussion in sec.~\ref{sec:dic}). The cross section for the
$\gamma p\rightarrow\gamma X$ scattering in LL accuracy is obtained by 
convolution of partonic cross sections with relevant LO parton densities.

\begin{figure*}
\vskip 0.5cm
\psfig{figure=dptjet.ps,height=7cm}
\vskip -7cm
\psfig{figure=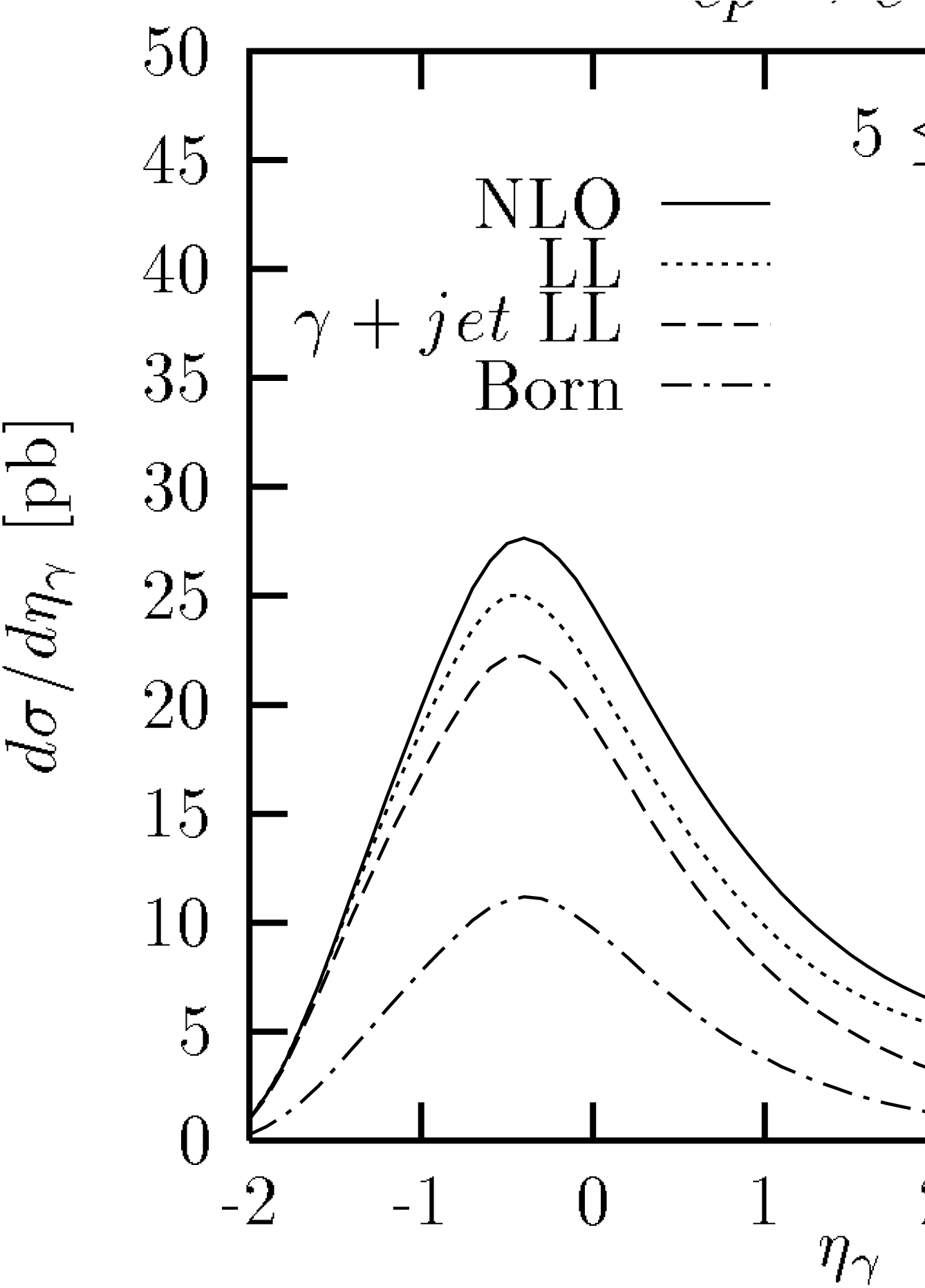,height=7cm}
\caption{The differential cross sections $d\sigma/d\eta_{\gamma}$ (a)
and $d\sigma/dp_T$ (b): NLO (solid line) and LL (dotted line) results
for isolated $\gamma$, 
and LL predictions for isolated $\gamma$ + $jet$ photoproduction
(dashed line);
the jet rapidity is assumed in the range: $-1.5\leq\eta_{jet}\leq 1.8$.
The dot-dashed lines show the Born contribution to the NLO cross section
for isolated $\gamma$.
GRV NLO (LO) parton densities in the proton and photon were applied in NLO
(LL) calculations.}
\label{fig:jet}
\end{figure*}

In fig.~\ref{fig:jet} we show the LL prediction for the isolated $\gamma$ 
photoproduction (dotted line) together with NLO predictions (solid line) 
and the Born contribution only (dot-dashed line).
The highest differences between the LL and NLO cross sections
are seen at the rapidity range $-0.5 < \eta_{\gamma} < 2.5$
where the LL results lie 10-20\% below the NLO ones (fig.~\ref{fig:jet}a).
For the $p_T$ distribution this difference is 10-14\% in the whole presented
range of the transverse momentum, $4\le p_T\le 20$ GeV (fig.~\ref{fig:jet}b).

We think that the observed difference between NLO and LL results together 
with the weak dependence on the $\bar{Q}$ scale discussed in 
sec.~\ref{results3} indicates reliability of the calculation.

\subsection{The comparison with other NLO results}\label{results5}

As we discussed in Sec.~\ref{sec:dic}, our NLO calculation of the DIC process 
differs from the ``$1/\alpha_s$''-type NLO analysis
presented in ref.~\cite{Gordon:1995km}$^-$\cite{Gordon:1998yt}, 
by set of diagrams included in the calculation.
We do not take into account $\alpha_s$ corrections to the single and double
resolved processes, which are beyond the NLO accuracy in our approach. 
On the other hand, we include the box diagram neglected
in~\cite{Gordon:1995km}$^-$\cite{Gordon:1998yt}. 
(The double resolved subprocesses are 
included in both analyses.) 

We compare our results and the results of the LG 
calculation~\cite{Gordon:1998yt} (using $N_f$=4 and $\bar{Q}=p_T$)
for the isolated final photon ($R$=1, $\epsilon$=0.1) in the 
kinematical range as in the ZEUS analysis~\cite{Breitweg:2000su}  
(i.e. for $-0.7\le\eta^{\gamma}\le 0.9$ and $0.2\le y\le 0.9$).
First we use the GRV photon parton densities.
The LG predictions for $d\sigma /dp_T$ cross section are about 20\% higher 
than ours in the presented range of transverse momentum, $4\le p_T\le 20$ GeV.
For $d\sigma /d\eta^{\gamma}$ cross section (with $5\le p_T\le 10$ GeV)
the biggest differences are at $\eta^{\gamma}$=0.9 where the LG results 
are about 35\% higher. The differences decrease towards negative rapidity 
values and are negligible at $-0.7\le\eta^{\gamma} < -0.5$.
For $y$ range limited to low values only, 0.2 $< y <$ 0.32, the LG cross 
section is higher than ours by up to 20\% at positive $\eta_{\gamma}$, 
while at negative $\eta_{\gamma}$ it is lower by up to 10\%.
For large $y$ values, 0.5 $< y <$ 0.9,
where our predictions agree with data, the LG results are higher than 
ours by up to 80\% (at $\eta_{\gamma}$ = 0.9).

As it was already discussed, for $\bar{Q}=p_T$
the GS photon distributions lead to results lower by 11-14\% than those 
obtained with GRV densities 
at rapidities between -1 and 1 (see sec.~\ref{results2}). 
In calculations presented in~\cite{Gordon:1995km}$^-$\cite{Gordon:1998yt}
this difference is even twice larger.

The LG predictions~\cite{Gordon:1998yt}
obtained using the GS parametrization lie up to 20\%
below ours (also based on GS distributions, with $\bar{Q} = p_T$ and
$N_f = 4$) at rapidities
$-0.7\le\eta_{\gamma}\le 0.2$, and they are higher than ours by up to 30\%
for $0.2\le\eta_{\gamma}\le 0.9$~\cite{Gordon:1998yt,Breitweg:2000su}.

\section{The LL prediction for $\gamma + jet$ photoproduction}\label{ll}

The ZEUS Collaboration has also analyzed the prompt photon photoproduction
in which in addition a hadron jet is 
measured~\cite{Breitweg:1997pa,unknown:1998uj,:2001aq}.
In fig.~\ref{fig:jet} we show the LL prediction for the 
isolated $\gamma + jet$ final state~\footnote{The NLO calculation 
for $\gamma + jet$ photoproduction will be discussed in our next paper.}
together with the predictions for the $\gamma$ alone.
The following jet rapidity and transverse momentum are assumed:
$-1.5\leq\eta_{jet}\leq 1.8$ and $p_T^{jet}>5$ GeV, respectively. 
This additional cuts for the final state imposed on jets
decrease the cross section,
especially at high rapidities. The LL predictions for the $\gamma +jet$
are lower than those for the $\gamma$ production by 5-10\% at negative
rapidities and by 10-80\% at positive rapidity values.
The difference between both LL results is about 10\% in wide range
of transverse momenta, $\sim 6\le p_T\le 20$ GeV, and only for
the lower $p_T$ region, $4\le p_T\le 6$ GeV, it is higher (13-23\%).

\section{Summary}\label{sec:sum} 
Results of the NLO calculation, with NNLO contributions from double 
resolved photon processes and box diagram, for the isolated $ \gamma$ 
production in the DIC process at HERA are presented~\footnote{Our
fortran code is available upon request from azem@fuw.edu.pl.}.
The role of the 
kinematical cuts used in the ZEUS measurement~\cite{Breitweg:2000su} 
are studied in detail.

The results obtained using GRV parametrizations agree with the data
in shape and normalization for $p_T$ distribution.
For $\eta^{\gamma}$ distribution a good description of the data
is obtained for $\eta^{\gamma}>0.1$, while for $\eta^{\gamma}<0.1$
the data usually lie above the predictions. 
This discrepancy arises mainly from the low $y$ region, $0.2\le y\le 0.32$.
The beyond NLO terms,
especially a box contribution, improve the description of the data. 

We have studied the theoretical uncertainty of results due to the choice of 
the renormalization/factorization scale: $\bar{Q} = p_T/2, p_T, 2p_T$.
At high rapidities $\eta_{\gamma}> 3$, where the cross section
is small, the uncertainty is 10-30\%.
In a wide range of rapidities, $-2\le\eta_{\gamma}\le 2$, the dependence
on the $\bar{Q}$ scale is small, below 6\%.
Since we include some NNLO diagrams in our NLO calculation,
this stability of the predictions versus the change of the scale is 
especially important. The week dependence on the $\bar{Q}$ scale,
and not large differences between LL and NLO predictions (below 20\%)
allows to conclude that theoretical uncertainties of our NLO calculations
for an isolated photon production in the DIC process at HERA 
are relatively small.

We compared our results with the LG ones based on different set of 
subprocesses.
The cross section $d\sigma/dp_T$ obtained by LG is about 20\% higher than 
ours (for GRV photonic parton distributions). For the cross section
$d\sigma/d\eta_{\gamma}$ this difference is up to 
35\% at $\eta_{\gamma} = 0.9$.
The highest differences are present for high $y$ values only,
$0.5 < y < 0.9$, where on the other hand our predictions are in
agreement with the data. At low $y$ range, $0.2 < y < 0.32$, differences
between both calculations are smaller and none of them
describe the data well for rapidities below 0.1.

\section*{Acknowledgements}
We would like to acknowledge
P. Bussey, Sung W. Lee and L.E. Gordon for important discussions.
\newline\newline
Supported in part by Polish State Committee for Scientific
Research, grants number 2P03B18410 (1998-1999),
2P03B11414 (2000) and 2P03B05119 (2000-2001), by Interdisciplinary
Centre for Mathematical and Computational Modelling, Warsaw University,
Grant No G16-10 (1999-2000), and by European Commission 50th framework
contract HPRN-CT-2000-00149.

\twocolumn[
\section*{Appendix}\label{sec:app}

Here we present formulae for the subtraction term in the cross section for the 
production of the isolated photon (see sec.~\ref{sec:iso})
in the $\gamma p\rightarrow\gamma X$ scattering. The corresponding expressions
for the $ep$ reaction can be obtained using the equivalent photon
approximation (sec.~\ref{sec:epa}).
The subtraction term is the cross section for subprocesses in which
the energy of hadrons inside the cone of radius $R$ around the final
photon is higher than the energy of the photon multiplied by a small parameter
$\epsilon$,
\bea
\sum_{h} E_h > \epsilon E_{\gamma} .
\label{eq:e}
\eea
The cone is defined in the rapidity and azimuthal angle plane:
\bea
\sqrt{(\eta_h-\eta_{\gamma})^2 + (\phi_h-\phi_{\gamma})^2} \le R .
\label{eq:r}
\eea

Below we use the variables $v$ and $w$ defined in the following way:
\bea
v=1+{\hat{t}\over \hat{s}} ~,~~~~w={-\hat{u}\over \hat{s}+\hat{t}},
\nonumber
\eea
where $\hat{s}$, $\hat{t}$ and $\hat{u}$ are the Mandelstam variables
for the partonic subprocess,
\bea
\hat{s}=y x_{\gamma} x S_{ep} ~, ~~\hat{t}=y x_{\gamma} z T_{ep} ~, 
~~~~\hat{u}=x z U_{ep}, \nonumber
\eea
and $S_{ep}$, $T_{ep}$ and $U_{ep}$ stand for the Mandelstam variables
for $ep\rightarrow e\gamma X$ reaction,
\bea
S_{ep} = 2 p_e p_p ~, ~~~~T_{ep} = - 2 p_e p_{\gamma} ~,
~~~~U_{ep} = - 2 p_p p_{\gamma}. \nonumber
\eea
The $p_e$, $p_p$ and $p_{\gamma}$ denote the initial electron and proton
four-momenta and the four-momentum of the final photon, respectively. 
The fractional momenta $y$, $x_{\gamma}$, $x$ and $z$ are defined in 
secs.~\ref{sec:cross} and~\ref{sec:epa}.
\newline

The subtraction cross section consists from two contributions which arise
from subprocesses involving fragmentation function and from $\alpha_s$ 
corrections to the Born process:

\bea
E_{\gamma}{d^3\sigma_{sub}^{\gamma p\rightarrow\gamma X}\over
d^3p_{\gamma}} = 
E_{\gamma}{d^3\sigma_{frag}^{\gamma p\rightarrow\gamma X}\over
d^3p_{\gamma}} +
E_{\gamma}{d^3\sigma_{cor}^{\gamma p\rightarrow\gamma X}\over
d^3p_{\gamma}},
\eea
where
\bea
E_{\gamma}{d^3\sigma_{frag}^{\gamma p\rightarrow\gamma X}\over
d^3p_{\gamma}} = 
\sum_{b=g,q,\bar{q}}\sum_{c=g,q,\bar{q}}\int^{1/(1+\epsilon)}_0 {dz\over z^2}
\int_0^1 dx f_{b/p}(x,\bar{Q}^2)
D_{\gamma /c}(z,\bar{Q}^2)
E_{\gamma}{d^3\sigma^{\gamma b\rightarrow cd}\over d^3p_{\gamma}}+
~~~~~~~~~~~~~~~~~~~~~~~~~
\nonumber \\
+\sum_{a=g,q,\bar{q}}\sum_{b=g,q,\bar{q}}\sum_{c=g,q,\bar{q}}
\int^{1/(1+\epsilon)}_0 {dz\over z^2}\int^1_0 dx_{\gamma}\int^1_0 dx 
f_{a/\gamma}(x_{\gamma},\bar{Q}^2) 
f_{b/p}(x,\bar{Q}^2)
D_{\gamma /c}(z,\bar{Q}^2)
E_{\gamma}{d^3\sigma^{ab\rightarrow cd}\over d^3p_{\gamma}}
\label{eq:frag}
\eea
and
\bea
E_{\gamma}{d^3\sigma_{cor}^{\gamma p\rightarrow\gamma X}\over
d^3p_{\gamma}} = 
\sum_{i=1}^{2 N_f}\int_{x_0}^1 dx 
\Theta \left( {v(1-w)\over 1-v+vw} - \epsilon \right)
\cdot ~~~~~~~~~~~~~~~~~~~~~~~~~~~~~~~~~~~~~~~~~~~~~~~~~~~~~~~~~~~~~~~~~~
~~~~~~~~~~~~~~~
\nonumber \\ \cdot
{\left[ f_{q_i/p}(x,\bar{Q}^2)
E_{\gamma}{d^3\sigma_{sub}^{\gamma q_i\rightarrow\gamma q_i + g}\over
d^3p_{\gamma}}+ 
f_{q_i/p}(x,\bar{Q}^2)
E_{\gamma}{d^3\sigma_{sub}^{\gamma q_i\rightarrow\gamma g + q_i}\over
d^3p_{\gamma}}+
f_{g/p}(x,\bar{Q}^2)
E_{\gamma}{d^3\sigma_{sub}^{\gamma g\rightarrow\gamma q_i + \bar{q}_i}\over
d^3p_{\gamma}} \right] }
\label{eq:cor}
\eea
]

\twocolumn[
with
\bea
x_0 = {-y T_{ep}\over y S_{ep} + U_{ep}}.
\nonumber
\eea

The contribution~(\ref{eq:frag}) comes from $2\rightarrow 2$ single
and double resolved subprocesses in which the final $c$-parton
decays into the photon. Here the calculations are standard, as for 
non-isolated photon case~\cite{Duke:1982bj}$^-$\cite{Gordon:1994sm}.
The condition (\ref{eq:e}) is included via the upper limit of 
the integration over $z$: $z\equiv E_{\gamma}/E_c < 1/(1+\epsilon)$
(the hadron remnant of the $c$-parton takes the energy 
$E_h=E_c-E_{\gamma}$, so $E_h=(1/z-1)E_{\gamma} > \epsilon E_{\gamma}$).
The lower limit of the integration over the fractional momenta $z$, 
$x_{\gamma}$ and $x$ is formally zero but in fact this zero-value is 
inaccessible due to the delta function, 
$\delta (y x_{\gamma} x S_{ep}+y x_{\gamma} z T_{ep}+x z U_{ep})$,
in the partonic cross sections for $2\rightarrow 2$ subprocesses.
\\

The contribution~(\ref{eq:cor}) 
describes the $\alpha_s$ corrections to the Born process.
The diagrams involving the virtual gluon exchange do not contribute to the
subtraction term, and only $2\rightarrow 3$ processes are included.
In these processes the photon and two partons are produced. One of
the partons enters the cone of radius R around the photon, and its
energy should be higher than the photon's energy multiplied by $\epsilon$.
To fulfill this condition 
the integration is performed 
over $x$ values for which $v(1-w)/(1-v+vw) > \epsilon$.
There are three types of such processes:\newline
$\bullet$ $\gamma q\rightarrow\gamma q + g$ (with a quark inside the cone),
\newline
$\bullet$ $\gamma q\rightarrow\gamma g + q$ (with a gluon inside the cone),
\newline
$\bullet$ $\gamma g\rightarrow\gamma q + \bar{q}$ 
(with a quark inside the cone).
\newline

Below we present our analytical results for 
the $\alpha_s$ corrections contributing to the subtraction term. 
The results are obtained with the assumption that the angle between
the photon and the parton inside the cone is small.
The quark masses are neglected. 
All collinear singularities are shifted into the fragmentation functions 
$D_{\gamma/c}$ (infrared singularities do not appear in this calculations).
The $p_T$ stands for the transverse momentum of the final photon.
The partonic cross sections in (\ref{eq:cor}) are given by following
expressions:

\bea
E_{\gamma}{d^3\sigma_{sub}^{\gamma q_i\rightarrow\gamma q_i + g}\over
d^3p_{\gamma}} = 
{4\over 3} {\alpha_{em}^2\alpha_s\over \pi\hat{s}^2} e_{q_i}^4
{\left[ {(1-v+vw)^2+(1-v)^2\over (1-v+vw)(1-v)} P(\bar{Q}^2) +
{(Rp_T)^2\over \hat{s}} {(vw)^3\over (1-v+vw)(1-v)^2} \right]} ~,
\eea

\bea
E_{\gamma}{d^3\sigma_{sub}^{\gamma q_i\rightarrow\gamma g + q_i}\over
d^3p_{\gamma}} = 
{4\over 3} {\alpha_{em}^2\alpha_s\over \pi\hat{s}^2} e_{q_i}^4
{(Rp_T)^2\over \hat{s}}
{(1-v) {[}(1-v+vw)^2+(vw)^2{]}\over 
(1-v+vw)^5v(1-w) (vw)^2}{[}1+(1-v+vw)^4+v^4(1-w)^4{]} ~,
\eea

\bea
E_{\gamma}{d^3\sigma_{sub}^{\gamma g\rightarrow\gamma q_i + \bar{q}_i}\over
d^3p_{\gamma}} =
{1\over 2} {\alpha_{em}^2\alpha_s\over \pi\hat{s}^2} e_{q_i}^4
{\left[ {(vw)^2+(1-v)^2\over vw(1-v)}P(\bar{Q}^2) +
{(Rp_T)^2\over \hat{s}} {(1-v+vw)^4\over (vw)^2(1-v)^2} \right]} ~,
\eea

where

\bea
P(\bar{Q}^2) = {1+v^2(1-w)^2\over (1-v+vw)^2} \ln
{\left( {R^2p_T^2v^2(1-w)^2\over \bar{Q}^2} \right) } +1 ~.
\eea

~\newline ~
]


\begin{table*}
\begin{center}
\caption{The cross section for non-isolated and isolated final photon, isolated
photon with $0.2\le y\le 0.9$, and isolated photon with $0.2\le y\le 0.9$ and
$-0.7\le\eta^{\gamma}\le 0.9$.}
\label{tab:tot}
\vspace{0.5cm}
\begin{tabular}{|c|c|c|c|c|c|c|c|} 
\hline 
\raisebox{0pt}[12pt][6pt]{[pb]} & 
\raisebox{0pt}[12pt][6pt]{total} & 
\raisebox{0pt}[12pt][6pt]{Born} &
\raisebox{0pt}[12pt][6pt]{\cal{O}($\alpha_S$)} & 
\raisebox{0pt}[12pt][6pt]{box} &
\raisebox{0pt}[12pt][6pt]{single resolved}&
\raisebox{0pt}[12pt][6pt]{single resolved}&
\raisebox{0pt}[12pt][6pt]{double resolved} \\
\raisebox{0pt}[0pt][6pt]{} & 
\raisebox{0pt}[0pt][6pt]{} & 
\raisebox{0pt}[0pt][6pt]{} &
\raisebox{0pt}[0pt][6pt]{} & 
\raisebox{0pt}[0pt][6pt]{} &
\raisebox{0pt}[0pt][6pt]{initial $\gamma$} &
\raisebox{0pt}[0pt][6pt]{final $\gamma$} &
\raisebox{0pt}[0pt][6pt]{} \\
\hline
\raisebox{0pt}[12pt][6pt]{non-isolated} & 
\raisebox{0pt}[12pt][6pt]{226.2} & 
\raisebox{0pt}[12pt][6pt]{82.1} & 
\raisebox{0pt}[12pt][6pt]{8.7} & 
\raisebox{0pt}[12pt][6pt]{13.9} & 
\raisebox{0pt}[12pt][6pt]{54.7} &
\raisebox{0pt}[12pt][6pt]{24.6} &
\raisebox{0pt}[12pt][6pt]{42.2}\\
\raisebox{0pt}[0pt][6pt]{} & 
\raisebox{0pt}[0pt][6pt]{} & 
\raisebox{0pt}[0pt][6pt]{(36.3\%)} & 
\raisebox{0pt}[0pt][6pt]{(3.8\%)} & 
\raisebox{0pt}[0pt][6pt]{(6.1\%)} & 
\raisebox{0pt}[0pt][6pt]{(24.2\%)} & 
\raisebox{0pt}[0pt][6pt]{(10.9\%)} &
\raisebox{0pt}[0pt][6pt]{(18.7\%)}\\
\hline
\raisebox{0pt}[12pt][6pt]{isolated} & 
\raisebox{0pt}[12pt][6pt]{180.4} & 
\raisebox{0pt}[12pt][6pt]{82.1} & 
\raisebox{0pt}[12pt][6pt]{15.2} & 
\raisebox{0pt}[12pt][6pt]{13.9} & 
\raisebox{0pt}[12pt][6pt]{54.7} &
\raisebox{0pt}[12pt][6pt]{5.12} &
\raisebox{0pt}[12pt][6pt]{9.37}\\
\raisebox{0pt}[0pt][6pt]{} & 
\raisebox{0pt}[0pt][6pt]{} & 
\raisebox{0pt}[0pt][6pt]{(45.5\%)} & 
\raisebox{0pt}[0pt][6pt]{(8.4\%)} & 
\raisebox{0pt}[0pt][6pt]{(7.7\%)} & 
\raisebox{0pt}[0pt][6pt]{(30.3\%)} &
\raisebox{0pt}[0pt][6pt]{(2.8\%)} &
\raisebox{0pt}[0pt][6pt]{(5.2\%)}\\
\hline
\raisebox{0pt}[12pt][6pt]{isolated} & 
\raisebox{0pt}[12pt][6pt]{72.33} & 
\raisebox{0pt}[12pt][6pt]{23.6} & 
\raisebox{0pt}[12pt][6pt]{6.33} & 
\raisebox{0pt}[12pt][6pt]{6.54} & 
\raisebox{0pt}[12pt][6pt]{28.2} &
\raisebox{0pt}[12pt][6pt]{2.34} &
\raisebox{0pt}[12pt][6pt]{5.29}\\
\raisebox{0pt}[0pt][6pt]{$y$ cut} & 
\raisebox{0pt}[0pt][6pt]{} & 
\raisebox{0pt}[0pt][6pt]{(32.6\%)} & 
\raisebox{0pt}[0pt][6pt]{(8.8\%)} & 
\raisebox{0pt}[0pt][6pt]{(9.0\%)} & 
\raisebox{0pt}[0pt][6pt]{(39.0\%)} &
\raisebox{0pt}[0pt][6pt]{(3.2\%)} &
\raisebox{0pt}[0pt][6pt]{(7.3\%)}\\
\hline
\raisebox{0pt}[12pt][6pt]{isolated} & 
\raisebox{0pt}[12pt][6pt]{35.36} & 
\raisebox{0pt}[12pt][6pt]{13.6} & 
\raisebox{0pt}[12pt][6pt]{3.32} & 
\raisebox{0pt}[12pt][6pt]{3.41} & 
\raisebox{0pt}[12pt][6pt]{11.9} &
\raisebox{0pt}[12pt][6pt]{1.21} &
\raisebox{0pt}[12pt][6pt]{1.92}\\
\raisebox{0pt}[0pt][6pt]{$y, \eta_{\gamma}$ cuts} & 
\raisebox{0pt}[0pt][6pt]{} & 
\raisebox{0pt}[0pt][6pt]{(38.5\%)} & 
\raisebox{0pt}[0pt][6pt]{(9.4\%)} & 
\raisebox{0pt}[0pt][6pt]{(9.6\%)} & 
\raisebox{0pt}[0pt][6pt]{(33.7\%)} &
\raisebox{0pt}[0pt][6pt]{(3.4\%)} &
\raisebox{0pt}[0pt][6pt]{(5.4\%)}\\
\hline
\end{tabular}
\end{center}
\end{table*}

\vspace*{-5cm}

~\newpage

\end{document}